\documentclass[pre,twocolumn,twoside,showpacs,superscriptaddress,floatfix,nofootinbib]{revtex4-1}

\usepackage{graphicx,amssymb,amsmath,epsfig,color,enumerate,bm}
\usepackage{epstopdf}
\usepackage{etoolbox} 

\makeatletter
\appto{\appendix}{%
  \@ifstar{\def\theequation@prefix{A.}}%
          {}%
}
\makeatother
\epsfclipon

\makeatletter
\appto{\appendix}{%
  \@ifstar{\def\theequation@prefix{A.}}%
          {}%
}
\makeatother

\begin{document}
\title{Stochastic threshold in cell size control}
\author{Liang Luo$^{\ddagger}$}
\email[Corresponding author.~]{luoliang@mail.hzau.edu.cn}
\affiliation{Department of Physics, Huazhong Agricultural University, Wuhan 430070, China}
\thanks{L. L. and Y. B. contributed equally to this work.}
\author{Yang Bai$^{\ddagger}$}
\author{Xiongfei Fu}
\email[Corresponding author.~]{xiongfei.fu@siat.ac.cn}
\affiliation{CAS Key Laboratory for Quantitative Engineering Biology, Guangdong Provincial Key Laboratory of Synthetic Genomics, Shenzhen Institute of Synthetic Biology, Shenzhen Institutes of Advanced Technology, Chinese Academy of Sciences, Shenzhen, 518055, China}

\begin{abstract}

Classic models of cell size control consider cells divide while reaching a threshold, e.g. size, age, or size extension. The molecular basis of the threshold involves multiple layers of regulation as well as gene noises. In this work, we study cell cycle as first-passage problem with stochastic threshold and discover such stochasticity affects the inter-division statistics, which bewilders the criteria to distinguish the types of size control models. The analytic results show the autocorrelation in the threshold can drive a sizer model to the adder-like and even timer-like inter-division statistics, which is supported by simulations. Following the picture that the autocorrelation in the threshold can propagate to the inter-division statistics, we further show that the adder model can be driven to the timer-like one by positive autocorrelated threshold, and even to the sizer-like one when the threshold is negatively autocorrelated. This work highlights the importance to examine gene noise in size control.

\end{abstract}

\maketitle

\section{Introduction}
\begin{figure}
\centering
\includegraphics[width=8.6cm]{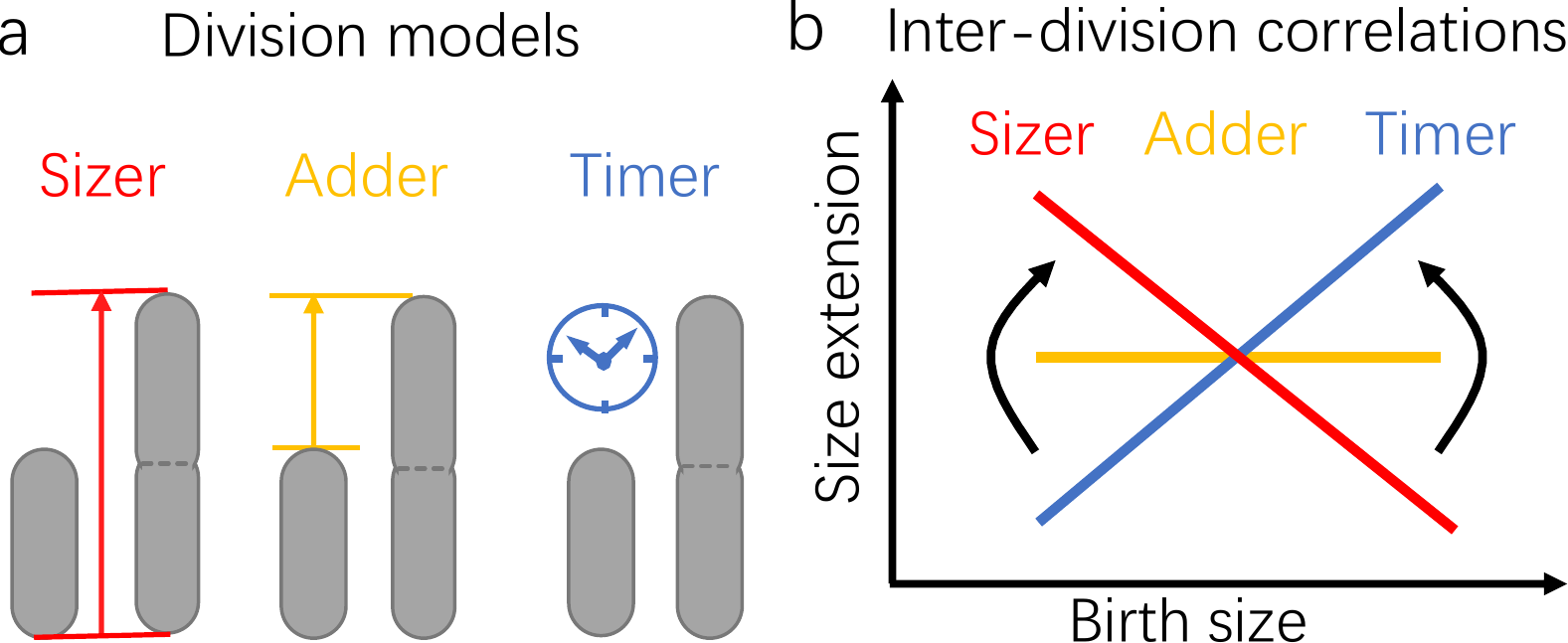}
\caption{\label{fig1} The inter-division correlation observed in experiments reflects both the regulatory mechanics and the correlation in the stochastic threshold. The three mechanics with cell size, size extension and cell age as cell cycle indicator ($x(t)$) (see (a)) led to the native inter-division correlation as sizer, adder, and timer (see (b)). The autocorrelation of the threshold $s(t)$ could drive the inter-division correlation either from sizer to adder to timer (with positive autocorrelation), or reverse (with negative autocorrelation). }
\end{figure}

Cell size homeostasis requires microbes to tightly control the fluctuations in exponential size growth and cell division \cite{willis17,suckjoon18}. The mechanism of the cell size control has been a puzzle for half a century since the discovery of cell growth law by Schaechter, Maal{\o}e, and Kjeldgaard\cite{schaechter58}. The modern experiments integrating the techniques of microfluidics and advanced imaging analysis shed a light on the puzzle, which allows direct measurements of cell cycles in a branch of the lineage\cite{suckjoon10}. The size control dynamics have been investigated since then in the phenomenological styles\cite{amir14, marco17,you15,brenner18}, focusing on the over-generation series of birth size $x_b$, division size $x_d$ and inter-division time $\tau$. According to the over-generation correlations, the experimental data are classified into three types as: the sizer that the division size $x_d$ is independent of the birth size $x_b$, the adder that the size extension $\Delta =x_d-x_b$ is independent of $x_b$, and the timer that $\tau$ is independent of $x_b$. Concerning on the underlying mechanisms of the classes, three types of models have been proposed\cite{willis17,robert15,mckinney20,wolde21}, assuming cell division happens when certain accumulating division indicator reaches a characteristic threshold. Different indicators are assumed for different types of over-generation correlation, which could be the cell size for sizer, the size extension since birth for adder, and the cell age since birth for timer\cite{amir14,voorn93,suckjoon15,suckjoon17}, where the simplified principle can be hence summarized by Fig.\ref{fig1}.

The over-generation correlations have been important criteria for biologists to search for the signal molecule that regulates cell division\cite{willis17}. Since the adder correlation is widely reported by experiments for bacteria such as E. coli. \cite{wagner14,suckjoon15,willis17,witz19,paulsson21}, the indicator of size extension has been intensively investigated\cite{amir15,zheng16,amir17,witz19,suckjoon19,zheng20,marco21,marco21b}. An attractive mechanism arises from the previous studies that the formation of the division ring plays a key role communicating cell mass accumulation and cell division. The accumulation of the related FtsZ protein then becomes a strong candidate as the indicator of size extension \cite{witz19,suckjoon19,zheng20,marco21b}. The accumulation mechanics of adder mechanics hence draws attentions from the theoretical side\cite{singh16,pandey20,nieto20,jia21,luo21,lileilei21}, while the experiments searching the molecular mechanics of cell size control are still in progress.

Recent experiments reported conflicting data that the statistics shifts from the adder-like correlation to the sizer-like one in the slow growth condition\cite{you15,elf16,nieto20}. The correlation as a mixture of adder and sizer seems demanding more complicated regulatory mechanisms. There have been candidates. The mixed models have been proposed that the DNA replication initiates with sizer or adder control mechanism while the replication requires roughly constant time\cite{amir15,amir17,elf16,witz19}. The concurrent models states the cell divides only when both indicators reaches the threshold\cite{marco18,marco21,marco21b}. The models based on the dynamics of bio-chemical reactions seems also work\cite{brenner18,pandey20,lileilei21}. All the the models respect the principle on the correspondence between the indicator and the over-generation correlation. It is, however, not necessarily true. In the recent study by Berger and ten Wolde\cite{wolde21b}, it is discovered that a sizer model can also have adder-like over-generation correlation in the case of fluctuating threshold.

We realize the molecular determinants of division threshold would be involved in complex regulatory networks, e.g. DNA replication, divisome formation, or mass accumulation\cite{elf16,suckjoon19,zheng20}. Imposed to noises in the regulatory networks, the division threshold would surely follow certain stochastic process, of which not only the magnitude of fluctuation but also the autocorrelation matters.
The autocorrelation in the stochastic threshold may thus propagate into the inter-division correlation, which breaks the principle on the correspondence between the indicator and the over-generation correlation based on the un-correlated assumption.
This manuscript aims to clearly demonstrate that the hidden correlation in the stochastic threshold would lead to the significant shift of the observed inter-division correlation, which can dramatically differ from the native one.

In this work, we developed a framework by describing the cell division process as a first-passage process (FPP)\cite{redner}. We analytically demonstrate that the mechanics with cell size as division indicator would also lead to the adder-like and even timer-like over-generation correlation, due to auto-correlated stochastic threshold $s(t)$. In the case of limited statistics, as usual in experiment, the adder-like state can be hardly distinguished from that from the native mechanics with size extension as the indicator. We then demonstrate the mechanics regulating the added size would be driven to the timer-like one by positive correlated $s(t)$, and even back to the sizer-like one by the negative correlated $s(t)$. It allows a continuous shift from adder-like to sizer-like correlation in the slow growth conditions, which has been reported by recent experiments. A comprehensive picture is hence illustrated how the autocorrelation in threshold modifies the over-generation correlation type.

\section{First-passage framework for cell cycles}

\begin{figure}
\centering
\includegraphics[width=8.6cm]{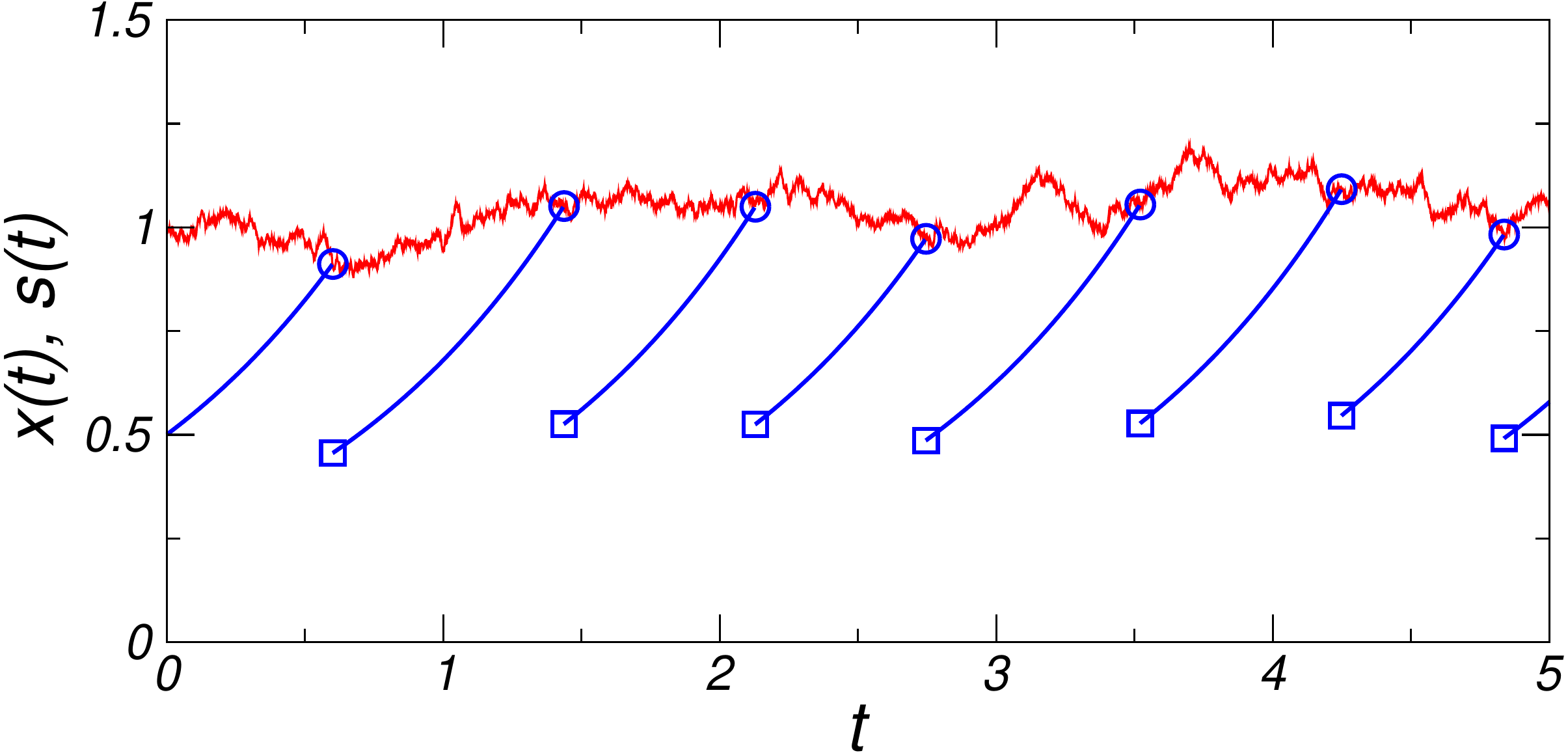}
\caption{\label{fig_traj} The typical trajectory of the stochastic sizer model. The red line shows the time series of size threshold, $s(t)$, following the Ornstein-Uhlenbeck process. The blue line shows the exponential growth of cell size, $x(t)$. The birth events are marked by the squares. The division events happen when the cell size reach the fluctuating threshold, marked by the circles. The parameters are set as the growth rate $\lambda=1$, the mean division size $s_0=1$, the feedback control strength $\gamma=1$ and the noise level  $\sigma=0.1$. }
\end{figure}

Let us consider a cell cycle since the birth at $t_0$ with the birth size $x_b$. The cell size grows exponentially with the fixed growth rate $\lambda$ as
\begin{equation}
\label{eq_growth}
x(t)=x_b\exp\left[\lambda (t-t_0)\right].
\end{equation}
The size control mechanics assumes cell division when some accumulation index reaches the division threshold. For simplicity of illustration, we take the sizer mechanics as an example first. It can be easily generalized to the other mechanics, such as the adder case studied in the later part of the paper.
The sizer mechanism suggests the cell divides when its size reaches the threshold $s$.  In the deterministic version, it leads to the fixed division size $x_d=s$, independent of the birth size $x_b$.  The randomness is introduced into the dynamics via the size threshold $s(t)$, which is assumed as a stochastic process controlled by certain feedback circuits around the mean value $s_0$. Assuming deterministic growth, the cell cycles are subordinated by $s(t)$, as shown in Fig.\ref{fig_traj}.

The cell divides when the growing $x(t)$ reaches the fluctuating $s(t)$ for the first time.
The cell cycle is equivalent to the FPP of the stochastic $s(t)$ to a shifting absorbing boundary at $x(t)$.
The survival probability that the cell has not divided till time $t$ can be expressed as
\begin{equation}
\label{eq_survival0}
S\left(t\vert s(t_0),t_0\right)=\int_{x(t)}^{\infty}ds\;P\left(s(t),t\vert s(t_0),t_0\right),
\end{equation}
where $P\left(s(t),t\vert s(t_0),t_0\right)$ is the probability density of $s(t)$ conditioned by known $s(t_0)$, or say, the packet of the probability density. In general, it deforms around the absorbing boundary. Exact solving for FPP is hence not trivial in most cases. (See e.g. \cite{ricciardi88} for Ornstein-Uhlenbeck process to a fixed absorbing boundary. )
In the case that the deformation is much slower than the absorbing process, the adiabatic approximation may help . Under this assumption, $P\left(s(t),t\vert s(t_0),t_0\right)$ can be approximated by the free propagator $G\left(s(t),t\vert s(t_0),t_0\right)$, which is, to be explicit, the probability density of $s(t)$ conditioned by known $s(t_0)$, without any concern of the boundary.
The first passage time can be then evaluated as
\begin{equation}
F\left(t\vert s(t_0),t_0\right)=-\frac{\partial S\left(t\vert s(t_0),t_0\right)}{\partial t}.
\end{equation}
The distribution of inter-division time follows
\begin{equation}
\label{eq_ptau}
P\left(\tau\vert x_b\right)=F\left(t=\tau+t_0\vert s(t_0)=2x_b, t_0\right),
\end{equation}
where the even division, $x_b=s(t_0)/2$, is assumed.
Noting Eq.(\ref{eq_growth}) and $P(x_d=x\vert x_b)dx=P(\tau\vert x_b)d\tau$,
the division size distribution can be expressed as
\begin{equation}
\label{eq_pxd}
P(x_d\vert x_b)=\frac{1}{\lambda x_d}P\left(\tau=\frac{1}{\lambda}\ln \frac{x_d}{x_b}\left\vert\right. x_b\right),
\end{equation}
which gives the full information of the correlation between the birth size and the division size. The added size distribution can be written as
\begin{equation}
P(\Delta\vert x_b)=P(x_d=x_b+\Delta\vert x_b).
\end{equation}
The joint probability would be convenient in comparison with the experimental data, which avoids the issue of  insufficient sampling on extreme $x_b$. In the case of known steady distribution $P(x_b)$, it can be written as
\begin{equation}
P(\Delta,x_b)=P(\Delta\vert x_b)P(x_b).
\end{equation}
The above framework allows us to estimate the concerned distributions from the Green's function $G\left(s(t),t\vert s(t_0),t_0\right)$ and the steady distribution $P(x_b)$.

The noise-free limit with the fixed size threshold $s(t)=s_0$ can be immediately solved as a simple example. In this  case,
\begin{equation}
G\left(s(t),t\vert s(t_0),t_0\right)=\delta\left(s(t)-s_0\right),
\end{equation}
where the Dirac delta is used.
The survival probability
\begin{eqnarray}
S\left(t \vert s(t_0)=s_0,t_0\right)&=&\int_{\frac{s_0}{2} e^{\lambda(t-t_0)}}^{\infty}ds\;\delta(s-s_0),\nonumber\\
	&=&H\left(\frac{1}{\lambda}\ln 2-(t-t_0)\right),
\end{eqnarray}
where $H(x)$ is the Heaviside step function.
It leads to
\begin{equation}
P(\tau\vert x_b)=\delta\left(\tau-\tau_c\right)
\end{equation}
with the expected inter-division time $\tau_c=\ln 2/\lambda$.
The division size distribution follows as
\begin{equation}
P(x_d\vert x_b)=\frac{1}{\lambda x_d}\delta\left(\frac{1}{\lambda}\ln\frac{x_d}{2x_b}\right)=\delta(x_d-s_0),
\end{equation}
where $x_b=s_0/2$ is applied. It is the typical behavior of a deterministic sizer.

\section{Sizer mechanics with stochastic threshold}

The stochastic size threshold $s(t)$ is in general controlled by certain feedback circuits around the mean value $s_0$. For the simplicity of analytic treatment, we consider the case of an Ornstein-Uhlenbeck (OU) process, which follows the Langevin dynamics by
\begin{equation}
\label{eq_OU}
ds/dt=-\gamma (s-s_0 )+\eta
\end{equation}
with uncorrelated Gaussian distributed noise $\left<\eta(t)\right>=0$ and $\left< \eta(t)\eta(t')\right>=2D\delta(t-t')$.
In the long time limit, the stationary distribution follows the Gaussian style as
\begin{equation}
P_{st}(\tilde{s})=\sqrt{\frac{1}{2\pi \sigma^2}}\exp\left[-\frac{\tilde{s}^2}{2\sigma^2}\right],
\end{equation}
where $\tilde{s}=s/s_0-1$, and the variance is controlled by the diffusion coefficient $D$ as $\sigma^2=D/(\gamma s_0^2)$. $\gamma$ reflects the strength of the feedback control on $s$ to the mean value $s_0$ which determines the typical correlation time as $t_c =1/\gamma$. Introducing the rescaled time $\tilde{t}=t/t_c$, the Green's function of the Ornstein-Uhlenbeck process\cite{risken} can be written as
\begin{equation}
G(\tilde{s},\tilde{t}\vert\tilde{s}',\tilde{t}')=\left[\frac{1}{2\pi \sigma^2\left(1-e^{-2(\tilde{t}-\tilde{t}')}\right)}\right]^{\frac{1}{2}}\exp\left[-\frac{\left(\tilde{s}-\tilde{s}'e^{-(\tilde{t}-\tilde{t}')}\right)^2}{2\sigma^2\left(1-e^{-2(\tilde{t}-\tilde{t}')}\right)}\right].
\end{equation}
One can see the OU process is positively correlated in the time scale of $t_c$.
When the generation time $\tau$ is also in this scale, the correlation can propagate into the inter-division size correlation, which eventually change the correlation classes.

\begin{figure}
\centering
\includegraphics[width=8.6cm]{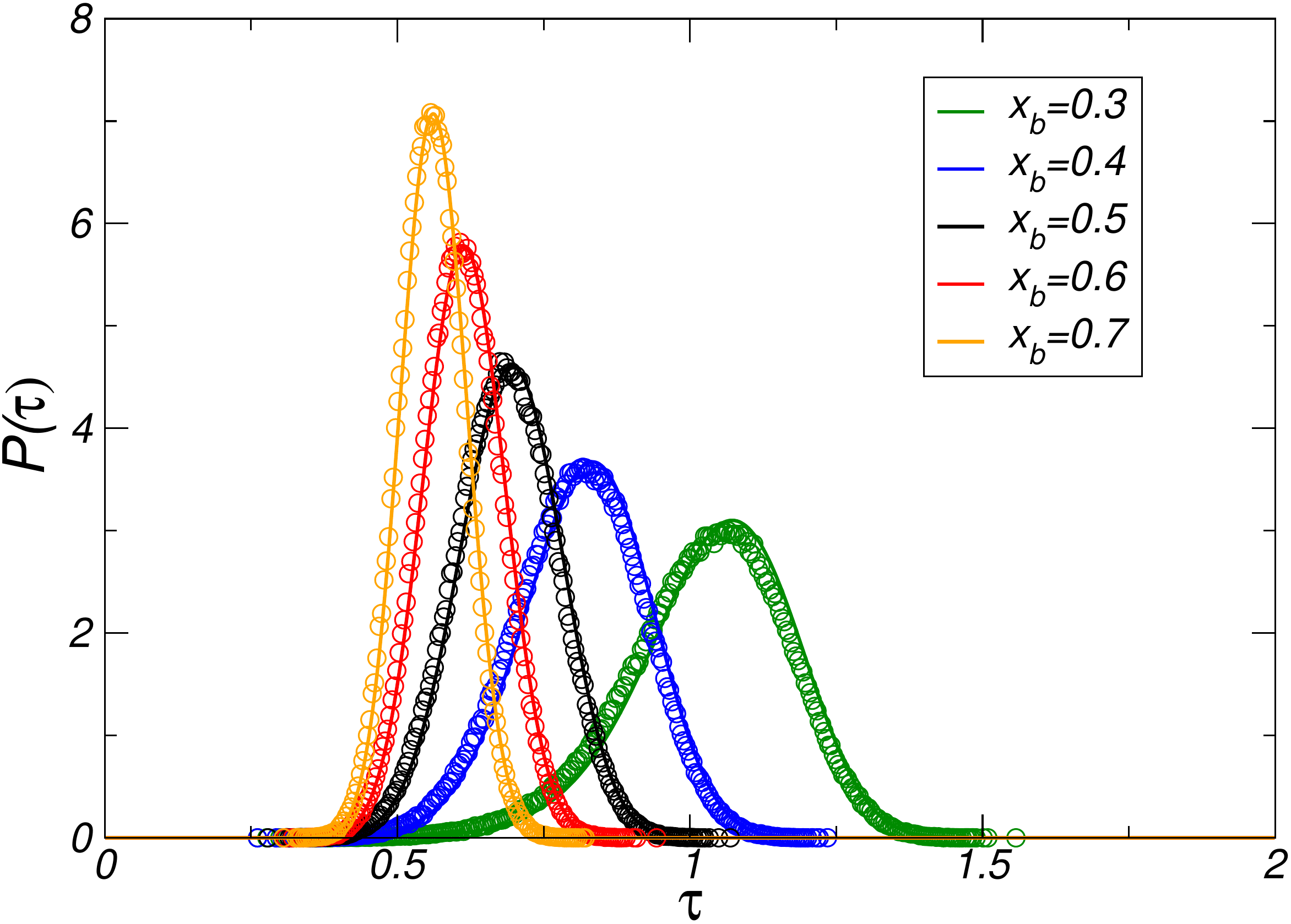}
\caption{\label{fig_ptau}The distribution of interdivision time $P(\tilde{\tau}\vert x_b)$ for the Ornstein-Uhlenbeck case with $\epsilon=1$, $\sigma=0.1$ and various $\tilde{x}_b$. The solid lines are given by Eq.(\ref{eq_ptauou}) and the symbols are from simulations.
}
\end{figure}

Since the cell divides when the growing $x(t)$ reaches the fluctuating threshold $s(t)$ for the first time. The cell cycle is equivalent to the first-passage process (FPP) of the fluctuating $\tilde{s}$ to a shifting absorbing boundary at $\tilde{x}-1$, where the size is rescaled by $\tilde{x}=x/s_0$. Adopting the adiabatic approximation, Eq.(\ref{eq_survival0}) - Eq.(\ref{eq_ptau}) lead to the distribution of rescaled inter-division time as
\begin{widetext}
\begin{eqnarray}
\label{eq_ptauou}
P(\tilde{\tau}\vert \tilde{x}_b)&=&\left(\frac{1}{2\pi \sigma^2}\right)^{1/2}\left(1-e^{-2\tilde{\tau}}\right)^{-3/2} \left[\epsilon \tilde{x}_b e^{\epsilon\tilde{\tau}}\left(1-e^{-2\tilde{\tau}}\right)+ \left(1-\tilde{x}_b e^{\epsilon\tilde{\tau}}\right)e^{-2\tilde{\tau}}- \left(1-2\tilde{x}_b\right)e^{-\tilde{\tau}}\right]\nonumber\\
& & \times \exp\left[{-\frac{\left(\left(1-2\tilde{x}_b\right)e^{-\tilde{\tau}}-\left(1-\tilde{x}_b e^{\epsilon\tilde{\tau}}\right)\right)^2}{2\sigma^2\left(1-e^{-2\tilde{\tau}}\right)}}\right],
\end{eqnarray}
where $\tilde{x}_b=x_b/s_0$, and $\epsilon=\lambda/\gamma$ reflects the ratio of two key time scales, i.e. the expected inter-division time $\tau_c$ and the typical correlation time $t_c$. As shown in Fig.\ref{fig_ptau}, the distribution is well agreed by the simulation results. It confirms the availability of the adiabatic approximation for this first-passage problem.
One can further obtain the other concerned distributions. Let us consider the ratio $\alpha=\tilde{x}_d/\tilde{x}_b$, the distribution of which follows
\begin{eqnarray}
\label{eq_pad}
P(\alpha\vert \tilde{x}_b)&=&\left(\frac{1}{2\pi\sigma^2}\right)^{1/2}\frac{1}{\alpha\epsilon}\left(1-\alpha^{-2/\epsilon}\right)^{-3/2}\left[\epsilon\alpha \tilde{x}_b(1-\alpha^{-2/\epsilon})+(1-\alpha \tilde{x}_b)\alpha^{-2/\epsilon}-(1-2\tilde{x}_b)\alpha^{-1/\epsilon}\right]\nonumber\\
& &\times\exp\left[-\frac{\left(\alpha^{-1/\epsilon}(1-2\tilde{x}_b)-(1-\alpha \tilde{x}_b)\right)^2}{2\sigma^2(1-\alpha^{-2/\epsilon})}\right].
\end{eqnarray}
\end{widetext}

\begin{figure*}
\centering
\includegraphics[width=12.9cm]{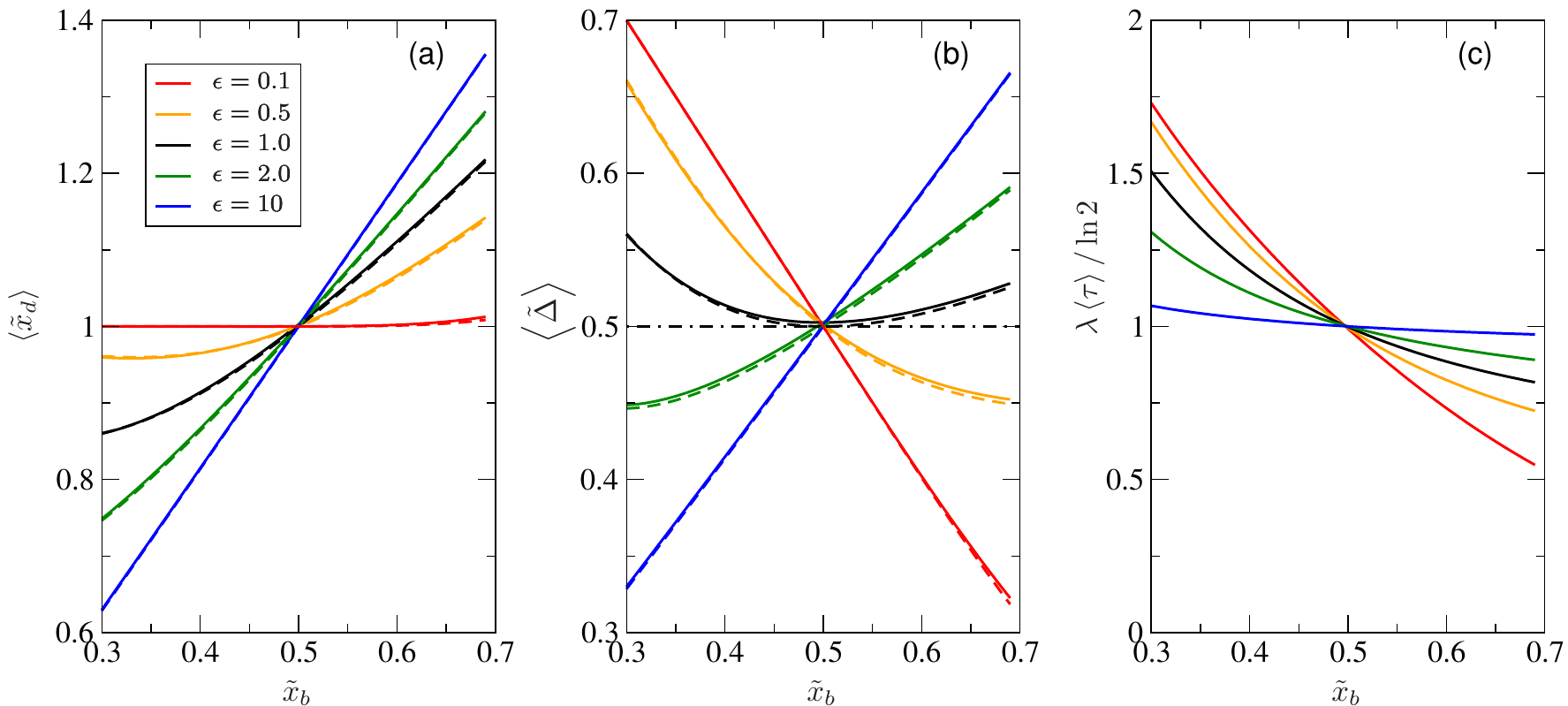}
\caption{\label{fig_xdmean} The inter-division correlation in the Ornstein-Uhlenbeck case. (a)The mean division size $\left<\tilde{x}_d\right>$ conditioned by the given birth size $\tilde{x}_b$ for $\sigma=0.1$ and various $\epsilon$. The dashed lines show the typical value $\hat{x}_d$ solved from Eq.(\ref{eq_peak}), which almost collapse to the mean values. (b) Same as (a), but for the mean added size $\left<\tilde{\Delta}\right>$. The dash-dotted line shows the perfect adder correlation for guidance. (c) The inter-division time $\left<\tau\right>$ rescaled by $\lambda/\ln 2$ for varying birth size $\tilde{x}_b$.  }
\end{figure*}

In the $\epsilon\ll 1$ limit, $s(t)$ and $s(t+\tau_c)$ has little correlation because $\tau_c\gg t_c$. The above distribution turns to be the Gaussian one as
\begin{equation}
\label{eq_pad1}
P_{\epsilon\rightarrow 0}(\alpha\vert\tilde{x}_b)=\frac{1}{\sqrt{2\pi\sigma^2}}\tilde{x}_b\exp\left[-\frac{(\alpha \tilde{x}_b -1)^2}{2\sigma^2}\right] .
\end{equation}
Noting $x_d=\alpha s_0\tilde{x}_b$, one can see the division size fluctuates around  $s_0$ with the distribution
\begin{equation}
\label{eq_pxd1}
P_{\epsilon\rightarrow 0}(x_d\vert x_b)=\frac{1}{\sqrt{2\pi\sigma^2 s_0^2}}\exp\left[-\frac{(x_d -s_0)^2}{2\sigma^2 s_0^2}\right].
\end{equation}
The variance is inherited from $s(t)$ as $\left< (x_d-s_0)^2\right>=D/\gamma$. It is the typical ``sizer'' behavior that the division size is governed by the threshold but independent of the birth size, as shown in Fig.\ref{fig_xdmean}(a). (See the line of $\epsilon=0.1$.)

In the $\epsilon\gg1$ limit, $s(t)$ and $s(t+\tau_c)$ are strongly correlated.
Eq.(\ref{eq_pad})  turns to be
\begin{equation}
\label{eq_pad2}
P_{\epsilon\rightarrow\infty}(\alpha\vert\tilde{x}_b)=\sqrt{\frac{\epsilon\tilde{x}_b^2}{4\pi\sigma^2}}\frac{2-\alpha+2\alpha\ln\alpha}{2\alpha(\ln\alpha)^{3/2}}\exp\left[-\frac{\epsilon\tilde{x}_b^2(\alpha -2)^2}{4\sigma^2\ln\alpha}\right].
\end{equation}
It is a distribution peaked around $\alpha=2$ with the variance propotional to $\sigma^2/\epsilon x_b^2$. The division size $x_d$ is always about twice of the birth size $x_b$.The inter-division time $\tau=\frac{1}{\lambda}\ln\alpha$ is largely independent of $x_b$, as shown in the right of Fig.\ref{fig_xdmean}(c). (See the line of $\epsilon=10$.) In simple words, it behaves as a typical ``timer''. The above two limit cases can be also plotted as a usual practice on the diagram of $\Delta=x_d-x_b$ versus $x_b$, according to two straight lines with slope $k=-1$ (sizer correlation) and $k=1$ (timer correlation), as shown by the $\epsilon=0.1$ case and the $\epsilon=10$ case in Fig.\ref{fig_xdmean}(b).

The crossover between the sizer and timer limits arises around $\epsilon\sim1$, which behaves like the adder as shown below. For general $\epsilon$, the full expressions of the distributions are rather complicated. The analytic estimation of the mean value is hard, if achievable. In the current case of small variance, one can turn to the mode of the distribution as an approximation.
Take the division size $\tilde{x}_d$ as the example, the distribution of which can be obtained from Eq.(\ref{eq_pad}) via the relation $\alpha=\tilde{x}_d/\tilde{x}_b$.
We note the peak of the distribution is governed by the factor
\begin{equation}
\label{eq_pxd}
P(\tilde{x}_d\vert \tilde{x}_b)\propto \exp\left(-z^2/(2g\sigma^2 )\right)
\end{equation}
where $g =(\tilde{x}_d/\tilde{x}_b)^{2/\epsilon}-1$ mainly modulates the peak width and $z=(2 \tilde{x}_b-1)+(\tilde{x}_d/\tilde{x}_b)^{1/\epsilon}(1-\tilde{x}_d)$. Setting $z=0$, one can estimate the position of the peak $\hat{x}_d$ by
\begin{equation}
\label{eq_peak}
(2\tilde{x}_b-1)+(\hat{x}_d/\tilde{x}_b )^{1/\epsilon}(1-\hat{x}_d)=0.
\end{equation}
In the $\epsilon\ll1$ limit, the above equation is satisfied only if $2\tilde{x}_b-1=0$ and $\hat{x}_d-1=0$, which gives the sizer behavior shown above. In the $\epsilon\gg1$ limit, Eq.(\ref{eq_peak}) requires $\hat{x}_d=2\tilde{x}_b$, which is the timer case discussed above. Concerning the relation between the typical added size $\hat{\Delta}\equiv\hat{x}_d-\tilde{x}_b$ and the birth size $\tilde{x}_b$, the crossover between the above two limits can be characterized by the slope $k\equiv d\hat{\Delta}/d\tilde{x}_b$.In the concerned regime around $\tilde{x}_b=1/2$, Eq.(\ref{eq_peak}) suggests
\begin{equation}
\label{eq_k}
k=2^{1-1/\epsilon}-1
\end{equation}
which smoothly shifts from $k=-1$ for the $\epsilon\ll1$ sizer limit to $k=1$ for the $\epsilon\gg1$ timer limit, as shown in Fig.\ref{fig_xdmean}(b). One may immediately note $k=0$ when $\epsilon=1$. In this case, the typical added size $\hat{\Delta}$ gently depends on the birth size and slightly deviates from the expected value $1/2$, as shown in Fig.\ref{fig_xdmean}(b). The model hence behaves like the adder one, bearing small errors.

\begin{figure}
\centering
\includegraphics[width=8.6cm]{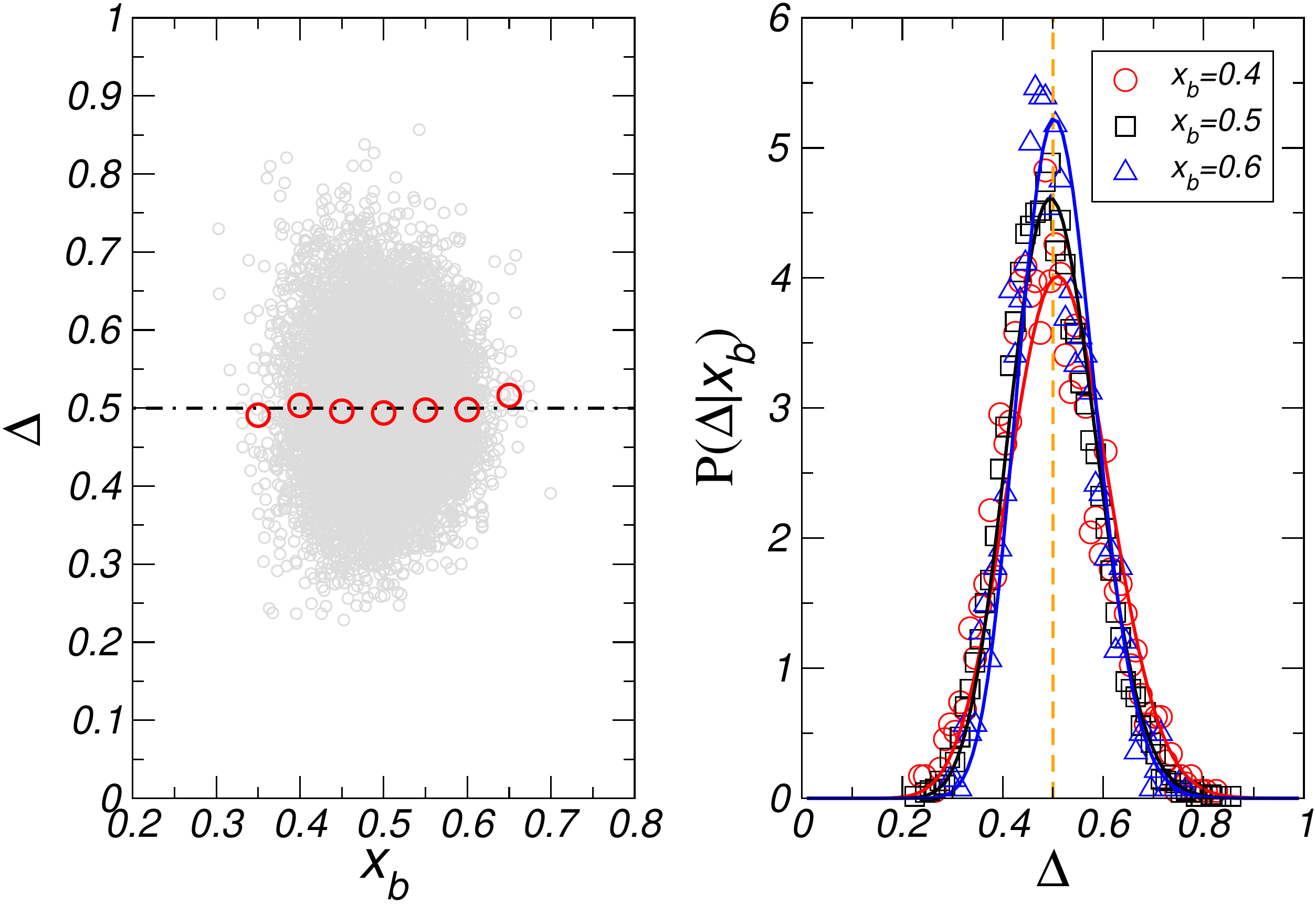}
\caption{\label{fig_pdelta} The statistics of simulation of $10^4$ cell cycles of sizer mechanics with the threshold $s(t)$ following OU process with $\epsilon=1$, $s_0=1$, and $\sigma=0.1$. (a) The scatter plot of the birth sizes versus the added sizes is shown by the gray circles. The mean added sizes for the given birth size are shown by the red circles. The black dash-dotted line indicates the added size for the perfect adder. (b) The distribution of the added size conditioned by various birth sizes, $P(\tilde{\Delta}\vert \tilde{x}_b)$ for $\epsilon=1$ and $\sigma=0.1$. The solid lines are the analytic results. The symbols are from simulation.}
\end{figure}

One may concern the deviation of the $\epsilon=1$ case from the perfect adder shown as the dash-dotted lines in Fig.\ref{fig_xdmean}(b) . The deviation can be, however, hardly identified in experiments, where the statistics is commonly limited. To illustrate this, we performed simulation with $10^4$  cell cycles. Figure \ref{fig_pdelta}(a) shows the $x_b-\Delta$ scatter plot, which looks just like that of the adder model. The mean added size slightly deviates for the extreme birth sizes. But the deviation might be ignored by eyes due to the poor statistics in these regions. The collapse of the added size distribution $P(\Delta\vert x_b)$ for various $x_b$ is another key observation in experiments \cite{suckjoon15}, which has been an important support to the accumulation mechanics \cite{singh16,luo21}. In the current model, the distribution slightly changes for various birth sizes, but look very similar in the range $x_b/s_0=0.4\sim0.6$, where one can merely observe insignificant differences of peak heights, as shown in Fig.\ref{fig_pdelta}(b).

All the above analysis bases on the model with $s(t)$ following OU process. To confirm the transition between the correlation types is induced by the autocorrelation in the threshold $s(t)$ but not else, we modified $s(t)$ from OU process to the Gaussian distributed random series. The positive autocorrelation is introduced into the series by the filter in Fourier space. The adder-like correlation arises again when the correlation time and the generation time matches. (See Appendix \ref{app_a} for details. )

In this section, we have demonstrated the positive autocorrelation in the threshold can propagate into the inter-division statistics, driving the sizer-type inter-division correlation between size extension and birth size to that of adder-like and even timer-like correlation. It can be the consequence of a more general scheme. The observables, such as the division size, are sampled from a hidden stochastic process $s(t)$ by the time interval $\tau_c$. The correlation of $s(t)$ may hence be inherited by the observables when $\tau_c$ is smaller than the correlation time of $s(t)$. This scheme can be generalized to the mechanism regulating other quantities, such as the added size shown in the next section.

\section{Adder mechanics: the effects of positive correlation and negative correlation}
\begin{figure*}
\centering
\includegraphics[width=12.9cm]{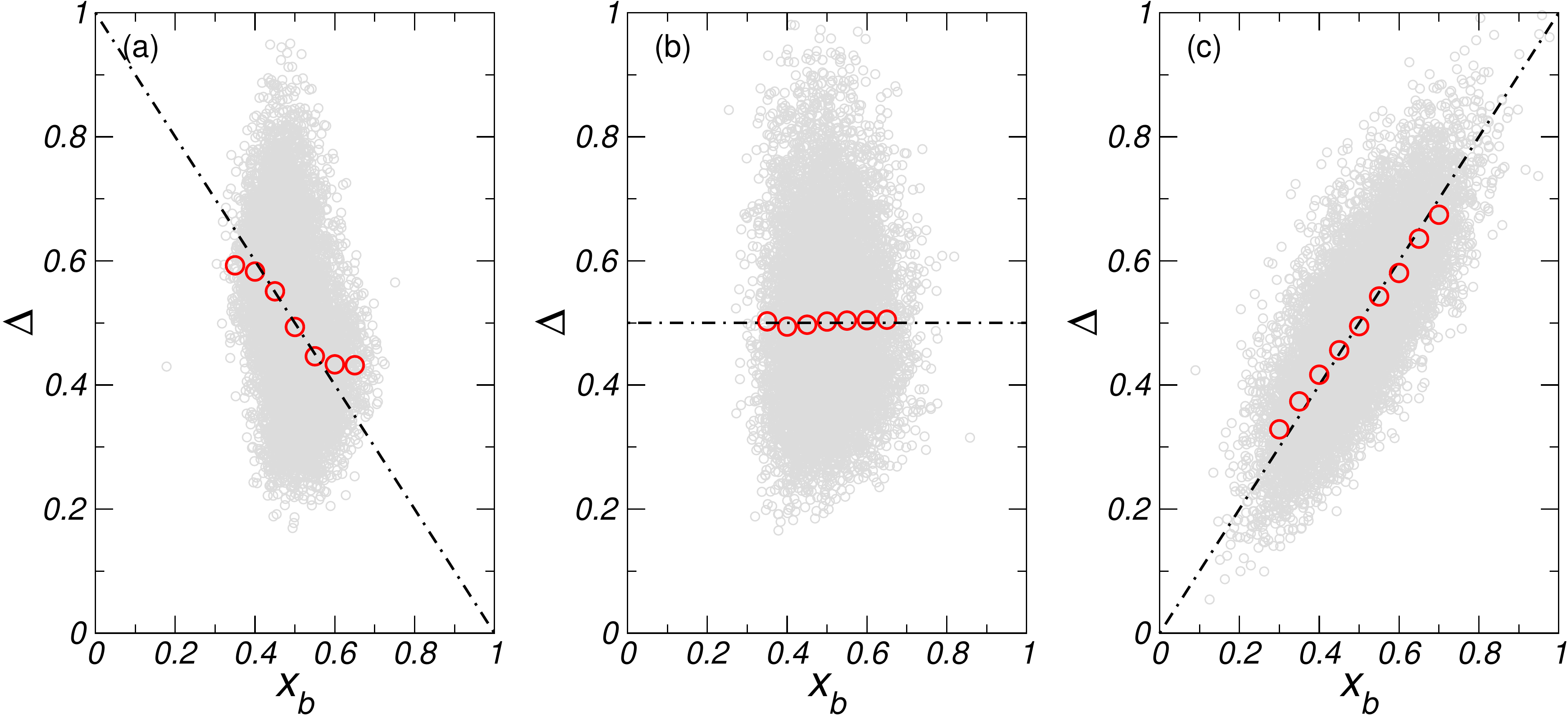}
\caption{\label{fig_deltamean_adder}The intrinsic correlation modifies the inter-division correlation of the accumulator mechanics, shown by the $x_b-\Delta$ scatter plot of simulation data (blue circles). The red circles denote the mean added size $\left< \Delta\right>$ for the given $x_b$. The dashed lines show the perfect sizer, adder, and timer correlations for guidance. (a) The oscillating threshold $s(t)$ drives the native adder correlation to the sizer-like one when the oscillation period $T\simeq2\tau_c$. (b) The native adder correlation of the accumulator with stochastic threshold $s(t)$ in the uncorrelated limit with $\epsilon=t_c\ln2/\tau_c=0.1$. (c) The positive correlated threshold $s(t)$ drives the native adder correlation to the timer-like one in the strong correlated limit with $\epsilon=10$.}
\end{figure*}

The dynamics with regulation on the added size can be achieved by the accumulator mechanics. It monitors an adder index $u$, which is reset to zero by birth and increases along with cell growth by $du(t)/dt=\lambda x(t)/\Delta_0$, where $x$ is the growing cell size, $\lambda$ is the growth rate. The cell division is considered as the first-passage of $u(t)$ to the adder threshold $s(t)$. In the deterministic limit with $s = 1$, one can easily see $x_d=x_b+\Delta_0$. It is the native adder correlation of the accumulator mechanics. In the stochastic version ignoring the correlation in noises, the adder correlation is kept as shown in Fig.\ref{fig_deltamean_adder}(b), as well known in literature. In presence of the correlation in threshold $s(t)$, the type of inter-division correlation may be modified either to time or to sizer.

\begin{figure}
\centering
\includegraphics[width=8.6cm]{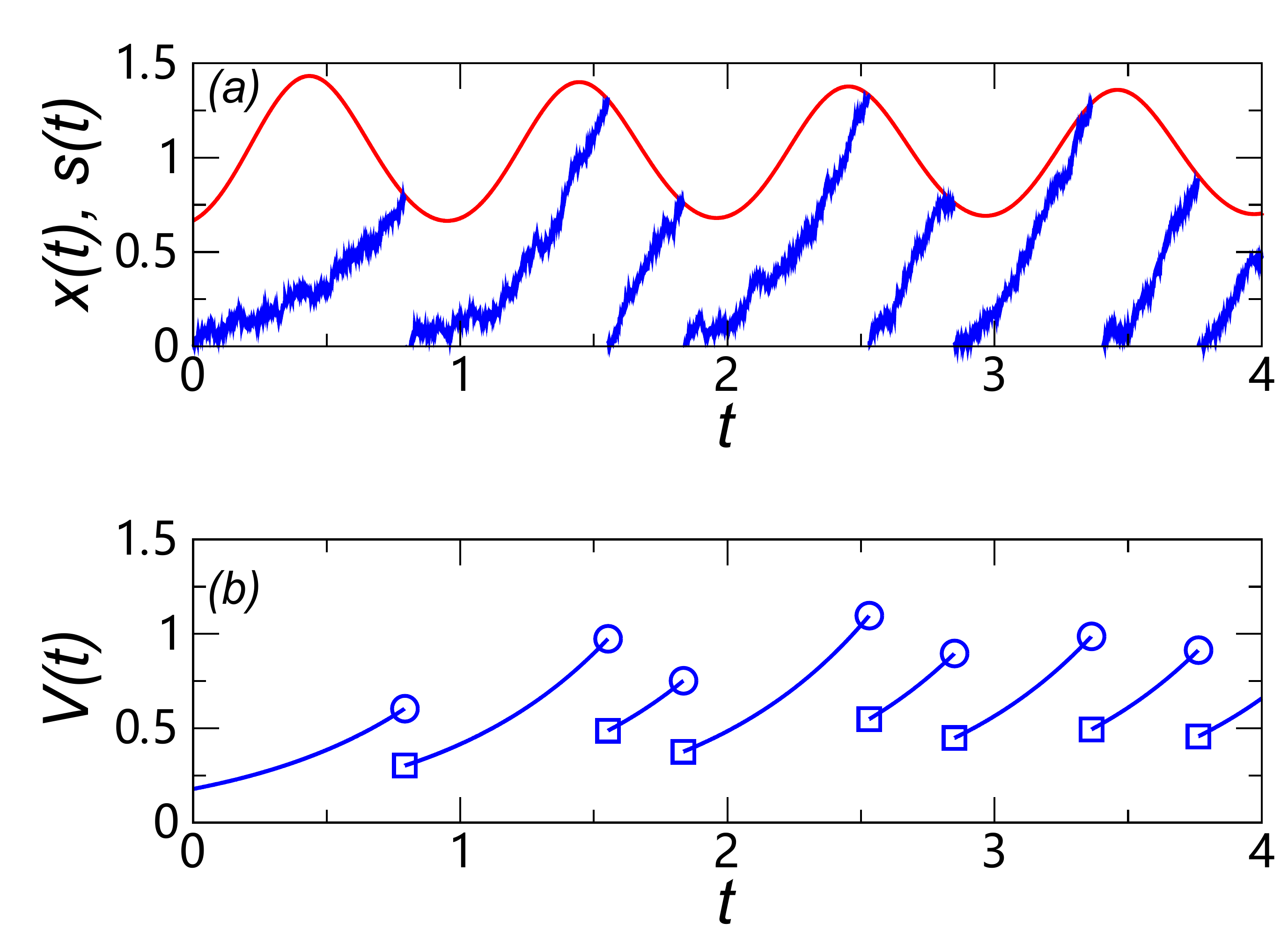}
\caption{\label{fig_trajosc} (a)The typical trajectory of the stochastic adder model with oscillating threshold. The red line shows the time series of adder threshold, $s(t)$, following the oscillating dynamics by \cite{elowitz00}. The blue line shows the accumulation of the adder index with noises, $x(t)$. (b) The cell size over generations. The birth events are marked by the squares. The circles denote the division events, which happen when the adder index reach the oscillating threshold shown in the upper figure. (c) The series of generation time, which is oscillating around the expected value due to the oscillating threshold. In the figures, the time is rescaled according to the oscillating period $T$. The expected generation is set as $\tau_c=T/2.2$, slightly smaller than the half period.  }
\end{figure}

To introduce the positive correlation, one can again assume $s(t)$ following the OU process with the intrinsic correlation time $t_c$. In the $t_c\gg \tau_c$ limit, $s(t)$ and $s(t+\tau_c)$ is strongly positive correlated. It modifies the type of inter-division correlation towards to more positive correlated case, i.e. the timer-like correlation, as shown in Fig.\ref{fig_deltamean_adder}(c). To drive an adder to the sizer-like one, the additional negative correlation is required.

The negative correlation can arise in the oscillating threshold $s(t)$, which may be a result of certain oscillatory gene network. The autocorrelation of $s(t)$ and $s(t+\tau)$ is negative when the periods of the oscillator $T\simeq2\tau_c$. To illustrate the adder to sizer transition, we performed the simulation of an accumulator with oscillating $s(t)$ following the dynamics by Elowitz and Leibler \cite{elowitz00}. Negative correlation in $s(t)$ is required to drive the adder mechanics to the sizer correlation. In this work, we has applied the oscillating dynamics of gene circuits by \cite{elowitz00} to demonstrate this transition. The dynamics considers the repressilator of three genes following
\begin{eqnarray}
\label{eq_osc}
\frac{dm_i}{dt}&=&-m_i+\frac{\alpha}{1+p_j^n}+\alpha_0\nonumber\\
\frac{dp_i}{dt}&=&-\beta(p_i-m_i),
\end{eqnarray}
where $m_i$ is the mRNA concentration of gene $i$, $p_i$ is the protein abundance in cell, $n$ is the Hill index, $i=1,2,3$, and $j=\mod(i,3)+1$. In the unstable regime of the parameter space, the dynamics oscillates. Supposing an adder mechanics with the threshold controlled by such oscillating circuits, negative correlation appears when the generation time is roughly half the period, as shown in Fig.\ref{fig_trajosc}(a). (Note the successive positions where the adder index $x(t)$ meets the threshold $s(t)$.)  The negative correlation propagates into the inter-generation size correlation via negatively regulated generation time, as shown in Fig.\ref{fig_trajosc}(c). The cell sizes (Fig.\ref{fig_trajosc}(b)) are hence more tightly controlled, leading to a sizer-like correlation shown in Fig.\ref{fig_deltamean_adder}(a). We note the negative correlation is a general feature of the oscillating $s(t)$ but not only for the dynamics defined by Eq.(\ref{eq_osc}). The above results hence stands for general oscillating thresholds.

\section{Discussions and summary}

The main results represented above guide us to a general picture of cell size control integrating randomness and correlation of stochastic process. It includes the naive deterministic model as the zero-noise limit, and the previously studied stochastic models as the uncorrelated limit. In the presence of gene noise and correlation, we show that the positive correlation of a stochastic threshold would drive the inter-division correlation observed in experiment shifting smoothly from sizer-like to adder-like and then to timer-like, while the negative correlation drives reversely, as indicated by the arrows in Fig.{\ref{fig1}}(b).

\begin{figure}
\centering
\includegraphics[width=6.4cm]{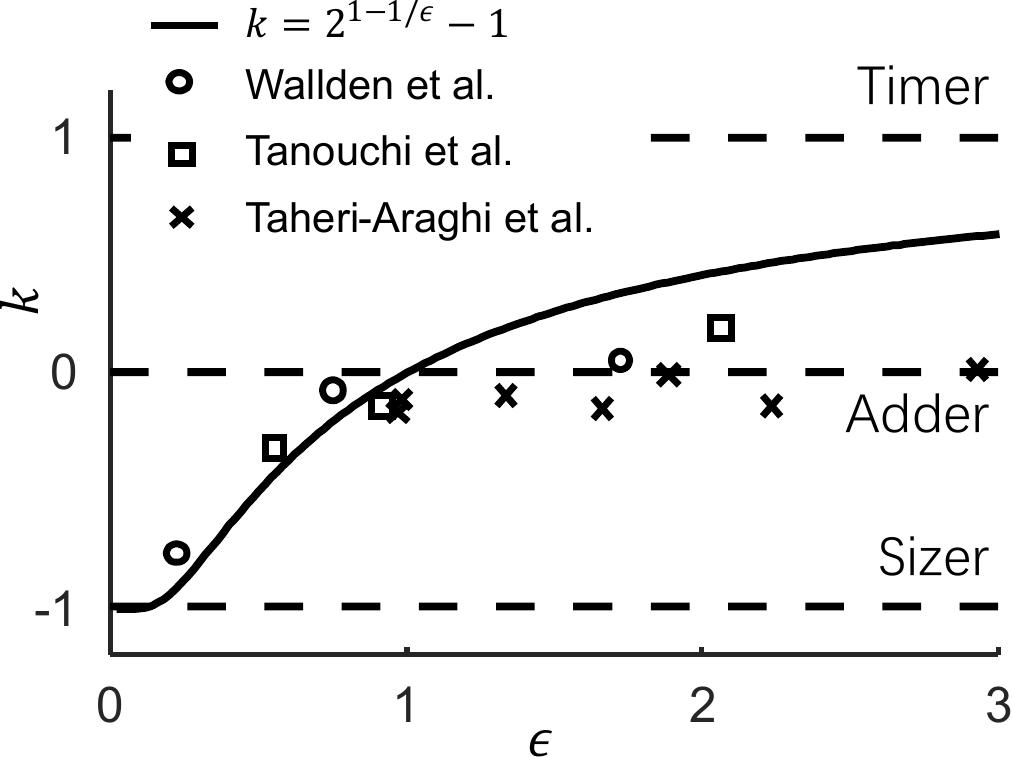}
\caption{\label{fig_exp}The autocorrelation in threshold s(t) modifies the inter-division correlation of the sizer mechanism. The slope $k=d\Delta/dx_b$ as a function of $\epsilon=\lambda/\gamma={t_cln 2/\tau}$ . The symbols are from the experimental data \cite{you15,suckjoon15,elf16}, assuming $t_c=1.2$hr. The black dashed lines show the perfect sizer adder timer correlations for guidance.}
\end{figure}

The analysis suggests the adder-like correlation appears when the correlation time of the process, $t_c$, matches the generation time $\tau$. When the two times differ, the correlation shifts to either the sizer or the timer one. The robust adder correlation is, however, observed in most experiments of fast growth conditions, as shown by the symbols in Fig.\ref{fig_exp}. The question hence arises how the two timescales would match in these experiments. A conjecture follows. Noise due to cell division may disturb the hidden process $s(t)$ and shorten its correlation to the generation time $\tau$. In simple words, $t_c$ is capped by $\tau$. The inter-division correlation is hence locked in the adder-like one in fast growth conditions. If this conjecture stands, one may expect in the slow growth case that $\tau$ would be much longer than the intrinsic correlation time $t_c$, the cell size control would slip from the adder to the sizer. It is surprising to us that the experiments \cite{you15,elf16} did observe the significant shift to sizer-like behavior in slow growth cases, as shown in the region of $\epsilon<1$ in Fig.\ref{fig_exp}.

Noting the shift between adder-like and sizer-like correlations, Tanouchi et al.\cite{you15} has proposed a phenomenological model  by assuming the relation between the birth size and division size $x_d=ax_b+b+\eta$, where $\eta$ is uncorrelated white noise. This kind of models \cite{you15,brenner18} decompose the randomness and the correlation into $\eta$ and the slope $a$, which help to capture the feature of experimental data. In the $x_b \text{ vs. }\Delta$ diagram, one can immediately read $a=k+1$, which slope $k$ is evaluated in this work by Eq.(\ref{eq_k}). We noticed that the deviation from the adder correlation has also been investigated in the framework of accumulation model since \cite{singh16}. It was suggested that in the case the division index is accumulated in the bursting style with the bursting size depending on the cell volume, the inter-generation correlation is more sizer-like. Nieto {\it et al.} \cite{nieto20} extended the accumulation model to the accumulation rate depending on the cell volume in non-linear way, which can also continuously bridge the sizer-like and adder-like behavior. In spite of the theoretical attempts,  the reason of the shift between adder and sizer is still unclear. It can be only answered by experiments offering more details, which may be a key to full understanding of bacterial cell size control mechanism.

The current study emphasizes the stochasticity in the circuit controlling cell division. Not only the magnitude of the threshold fluctuation but also the autocorrelation matters. It reminds us the studies on the connection between DNA replication and cell division. As a prominent example, the DNA replication that determines the later cell division are initiated when cells reach a critical threshold of active DnaA protein \cite{levin08}. Experimental evidences were reported for the possible regulations of active DnaA protein such as negative feedback of datA sites \cite{katayama12} or dnaA boxes \cite{hansen07}, which could generate autocorrelated stochasticity in the threshold. Berger and ten Wolde has proposed a cell cycle model controlling DNA replication initiation\cite{wolde21b}, which leads to their observation of the emergence of adder-like correlation from a control mechanics targeting on cell size but not the size extension.

In short summary, this study clearly shows how the autocorrelation of the intracellular process can modify the type inter-cycle correlation. The regulation mechanisms is hence not necessarily constrained by the inter-cycle correlation. It calls more careful inference on the experiment observations. We highlight that the simultaneous measurements of the inter-cycle correlation and the stochasticity of the intracellular variables are important to validate the cell cycle control models.

\begin{acknowledgements}
We thank C. Liu for discussion. This work is partially supported by the National Key Research and Development Program of China (2018YFA0903400, 2021YFA0910703), NSFC (11705064, 32071417), CAS Interdisciplinary Innovation Team (JCTD-2019-16), Guangdong Provincial Key Laboratory of Synthetic Genomics (2019B030301006), and Strategic Priority Research Program of Chinese Academy of Sciences (XDPB1803).
\end{acknowledgements}

\appendix
\section{Sizer mechanics with stochastic threshold: randomly-generated correlated time series}
\label{app_a}

\begin{figure}
\centering
\includegraphics[width=8.6cm]{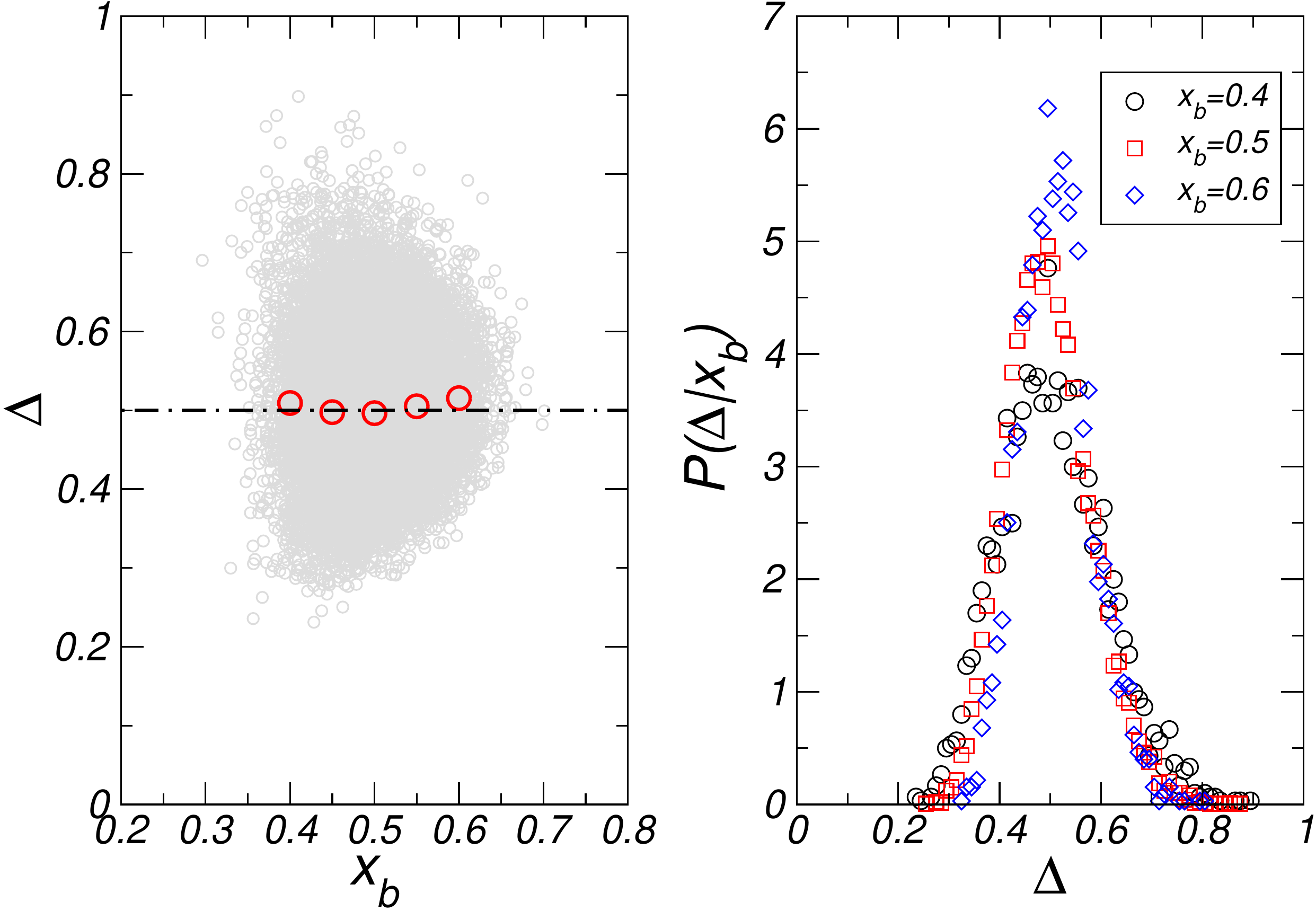}
\caption{\label{fig_deltacor} Same as Fig.3 in the main text, but with $s(t)$ generated by the filter in Fourier space. }
\end{figure}

To demonstrate the shift of the inter-generation correlation is purely induced by the autocorrelation in $s(t)$, we further test the dynamics on the generated the random series $s(t)$ with autocorrelations in the time scale $t_c$.
The random series is generated by three steps. First generate the random mode in Fourier space $z(k)=u(k)+i v(k)$, where $u$ and $v$ are uncorrelated normal distributed real number and $i=\sqrt{-1}$. Then introduce Gaussian filter in Fourier space by $\tilde{z}(k)\equiv z(k)\exp(-\sigma_f k^2)$. Finally, the inverse Fourier transform $\mathcal{F}^{-1}(\tilde{z})=r(t)+i s(t)$ offers two independent series $r(t)$ and $s(t)$, both correlated with itself in the time scale $t_c$, which is determined by $\sigma_f$.
Simulation has been performed for $2\times10^4$ cycles on the random $s(t)$. It shows that the inter-generation correlation is driven to the adder style when $t_c\ln2/\tau\simeq1$, as shown in Fig.\ref{fig_deltacor}.

\bibliography{sizer2adder}

\end{document}